\documentclass[twocolumn,showpacs,preprintnumbers,amsmath,amssymb,prb,superscriptaddress]{revtex4}

\usepackage{graphicx}
\usepackage{color}
\usepackage{pstricks}
\usepackage{epsfig}
\usepackage{overpic}

\begin{document}

\title{ Rydberg crystallization detection by statistical means }
\author{D.~Breyel}
\affiliation{Institut f\"ur Theoretische Physik, Universit\"at Heidelberg, Philosophenweg 19, D-69120 Heidelberg, Germany}
\author{T.~L.~Schmidt}
\affiliation{Department of Physics, Yale University, 217 Prospect Street, New Haven, CT 06520, USA}
\affiliation{Departement Physik, Universit\"at Basel, Klingelbergstrasse 82, 4056 Basel, Switzerland}
\author{A.~Komnik}
\affiliation{Institut f\"ur Theoretische Physik, Universit\"at Heidelberg, Philosophenweg 19, D-69120 Heidelberg, Germany}
\date{\today}



\begin{abstract}
 We investigate an ensemble of atoms which can be excited into a Rydberg state. Using a disordered quantum Ising model, we perform a numerical simulation of the
experimental
 procedure and calculate the probability distribution function $P(M)$ to create a certain number of Rydberg atoms $M$, as well as their pair correlation function.
Using the latter, we identify the critical interaction strength above which the system undergoes a phase transition to a Rydberg
 crystal. We then show that this phase transition can be detected using $P(M)$ alone.
\end{abstract}

\maketitle


\section{Introduction}
\label{sec:intro}

Recent experimental progress in producing and controlling highly excited atomic and molecular aggregates have triggered a number of experimental and theoretical
works on interacting Rydberg systems.\cite{PhysRevA.81.023406, PhysRevA.83.041802, PhysRevA.76.013413, PhysRevA.70.042703, PhysRevLett.107.060406,E.Urban2009,
PhysRevA.80.033422, PhysRevLett.100.113003, PhysRevLett.104.013001, PhysRevA.83.050501, PhysRevLett.100.243201, PhysRevA.73.062507, PhysRevLett.99.163601,
PhysRevLett.100.013002,2012arXiv1202.2779H, 2012arXiv1202.2012A}
One of the most prominent phenomena observable in such systems is the dipole blockade, \cite{PhysRevLett.93.063001,PhysRevLett.87.037901}
which is a consequence of effective interactions between atoms in Rydberg states with principal quantum numbers $n\sim 30-80$.\cite{PhysRevLett.93.163001}
The physical reason for the blockade can be summarized as follows: the large dipole moment of a Rydberg atom induces sizeable energy level shifts in the atoms in its
vicinity. As a result, atoms within a certain blockade radius $R_B$ cannot be excited, even though they are subjected to the same electromagnetic field as the
proper Rydberg atoms.

Typical experiments are conducted on atom ensembles at ultralow temperatures and probe these systems on timescales during which almost no particle movement is possible.
Therefore, the cloud of atoms can be considered as `frozen' in a more or less disordered constellation.\cite{PhysRevLett.80.253} After the required fine-tuned
electromagnetic fields are switched on, a number of atoms undergo a transition into the highly excited Rydberg states. If $R_B$ is larger than the average
interatomic distance, only a fraction of the atoms can be excited, while the rest remains in the ground state due to the blockade effect.

The spatial arrangement of the Rydberg atoms within the cloud has very interesting features. In Ref.~[\onlinecite{1367-2630-12-10-103044}], the concept of a Rydberg
crystal was put forward. The blockade region formed around an excited atom can be modeled as an effective repulsive interaction between the
Rydberg atoms. It might be responsible for an emergent long-range order of the Wigner crystal type\cite{PhysRev.46.1002}. However, its detection is extremely
difficult. The primary method for detecting long-range order is spectroscopy, which is difficult to reliably realize in experiments.\cite{2012arXiv1202.0699J} 
Other experiments benefit from
the low ionization energy of the Rydberg atoms by (pulsed) electric field ionization (cf. e.g. [\onlinecite{PhysRevA.18.229, PhysRevLett.99.163601,
PhysRevLett.98.023004, PhysRevLett.80.249, PhysRevLett.88.133004, PhysRevLett.93.153001, PhysRevLett.95.253002, PhysRevLett.100.243201, PhysRevLett.82.1839,
PhysRevLett.108.063008, PhysRevA.80.033422, PhysRevLett.80.253, PhysRevLett.95.233002, PhysRevLett.90.143002, 0022-3700-14-21-003, PhysRevLett.100.013002, PhysRevLett.104.173602,
PhysRevLett.107.103001, PhysRevLett.93.163001, PhysRevLett.93.063001, PhysRevLett.99.073002, PhysRevLett.97.083003}] ).

In this paper we propose a statistical method of detecting and analyzing the physical properties of the Rydberg crystallization phenomenon and discuss its predictive
power. The idea roots in the experimental procedure itself. A typical measurement cycle starts with the generation of an ultracold atomic cloud and the
subsequent excitation of a fraction of the atoms to a Rydberg state. Afterwards, measurements are performed, during which the Rydberg atoms are eventually de-excited.
Then, the system is ready for another preparation.\cite{1367-2630-10-4-045031, PhysRevLett.82.1839, PhysRevLett.95.233002, PhysRevLett.97.083003, Weidemuller2009}
An important point is that the experiments are performed with the same number of atoms and using the same electromagnetic fields in every cycle. However, since the
arrangement of atoms varies between cycles, it is necessary to calculate statistical averages of the observables of interest. The simplest observable is the number
of Rydberg atoms $M$ in the cloud. The fact that for a given $M$, a regular arrangement of the Rydberg atoms on a lattice minimizes their interaction energy, should
be visible in the probability distribution $P(M)$.
As we shall show below there are indeed differences between the histograms
for crystallized and random phases. However, the pair correlation function turns out to possess even higher predictive power.
\cite{2012PhRvL.108a3002G, PhysRevLett.107.103001}

This article is organized as follows.
In Sec.~\ref{sec:model} we shall introduce the model and connect its parameters to possible experimental setups. We shall define the relevant observables and explain
 the details of our numerical implementation. The results of the calculation together with a thorough analysis of the arising features is contained in
Sec.~\ref{sec:results}. Section \ref{sec:sumandcon} is devoted to the summary of results.


\section{The model and simulation method}
\label{sec:model}

We assume that in every measurement cycle the system consists of $N$ atoms located at randomly chosen positions ${\bf r}_i, i=1,\dots N$, uniformly distributed over
the entire system volume.
\footnote{This marks a difference to the recent publication [\onlinecite{1367-2630-12-10-103044}], which used a lattice.}
We focus on the case of a frozen Rydberg gas, where the kinetic energy is negligible. 
Each of the atoms can either be excited to a Rydberg state or stay in its ground state. Since the electrostatic properties do not depend on the details
of the Rydberg states, each atom can be modeled as a two-level system, and we describe the ensemble as a set of $N$ randomly arranged, interacting spin-$1/2$ systems.
Hence, the Hamiltonian reads\cite{PhysRevA.72.063403}
\begin{align}
H &= - \frac{\Delta}{2} \sum\limits_{i=1}^{N} \sigma_{z}^{(i)}
	      + \frac{\Omega}{2} \sum\limits_{i=1}^{N} \sigma_{x}^{(i)} \notag \\
	      &+ \frac{C}{4} \sum\limits_{i=2}^{N} \sum\limits_{j=1}^{i-1} \frac{(1 + \sigma_{z}^{(i)})(1 + \sigma_{z}^{(j)})}{|{\bf r}_{i}-{\bf r}_{j}|^{6}},
\label{eq:hamiltonian}
\end{align}
where $\sigma^{(i)}_{x,z}$ denote the Pauli matrices. This can be interpreted as a generalized spin-$1/2$ quantum Ising model. $\Omega$ is the frequency of the exciting
 laser and, in the spin language, represents a magnetic field perpendicular
to the quantization axis, which we chose to be the $z$-axis. The detuning, i.e., the difference between the laser frequency to the resonance frequency of the Rydberg
state, is denoted by $\Delta$. It
corresponds to a magnetic field applied in $z$-direction. The third parameter, $C$, indicates the strength of an effective interaction between excited
atoms and causes the dipole blockade explained above.\cite{Amthor2009} We note that we are only interested in the case $\Delta > 0$, because otherwise it is energetically
 not favorable to excite atoms. This allows us to use $\Delta$ as a basic energy scale.

An adequate modeling of the system also requires geometrical constraints describing the trap potentials. In the following we shall consider different scenarios:
(i) a 1D system with open boundaries,\footnote{We allow for interactions to act `through' the particles, that is why it is more appropriate to talk about quasi-1D
systems rather than 1D ones.} (ii) a 3D system with open boundaries, (iii) a 1D system with periodic boundary conditions. Options (i) and (ii) are very natural models
for realistic experiments, but make it difficult to extrapolate the presented numerical results towards realistic system sizes. Option (iii), on the other hand, is
perfectly
suitable for the calculation of correlation functions and is easily scalable to large system lengths. The crucial feature of our 1D model is a rather large ``coordination
number'', i.e., the number of atoms which interact significantly with any given Rydberg atom. This feature is shared by any generic 3D realization of the system
up to some irrelevant spatial distribution parameter. That is why we believe that the physics in 3D is expected to be very similar to our 1D model.

Just as in the actual experiments we calculate averages over the large number of different, randomly sampled atom arrangements. In every cycle
the effective model is a long-range quantum Ising model with a set of coupling constants generated by the atom positions ${\bf r}_i$. For every such constellation, we
determine the ground state and evaluate the number of Rydberg atoms $M$ (which is equivalent to
the magnetization in the Ising model), the density profile and correlation functions. The most difficult task is finding the ground state. We use one method,
numerical diagonalization of the Hamiltonian matrix, in three different variations. For small atom numbers ($N \approx 12$) the Hamiltonian
can be diagonalized exactly. For larger numbers of
atoms ($N \approx 30$), we truncate the Hilbert space in different ways in order to speed up the numerical diagonalization. On the one hand,
this truncation can be done by only keeping states in which the number of Rydberg atoms $M$ remains below a certain threshold $M^*$. A different approach is to only
use the basis states with the lowest energy expectation values to create an effective Hamiltonian, which is then diagonalized.
We checked the reliability of both procedures by changing the respective cutoff parameter.

To calculate the number of Rydberg atoms and the correlation function we proceed as follows. For a given random distribution of atoms, the Hamiltonian matrix
is expressed in a basis consisting of states in which an integer number of atoms is in the Rydberg state while the rest is in the ground state. The corresponding
Hilbert space is then truncated as explained above, and
%
%
the smallest eigenvalue and the corresponding eigenvector (the ground state $|GS\rangle$) are obtained numerically. The number of Rydberg atoms in the ground state is now found from
\begin{equation}
M = \sum\limits_{i=0}^{\tilde{N}} M(i) |v_{i}|^2,
\label{eq:magnetisation} \,
\end{equation}
where $\tilde{N}$ is the dimension of the truncated Hilbert space, $M(i) = \sum_{j=1}^N \langle i | (1 + \sigma_z^{(j)}) | i \rangle$ the number of Rydberg atoms in the basis state $|i\rangle$ and $v_{i} = \langle i | GS \rangle$ is the overlap between $|i\rangle$ and the ground state. One
easy way of numbering the basis states is to assign a ``1'' to a Rydberg atom and a ``0'' to a ground state atom. In this way, every basis state can be uniquely mapped to the binary representation of a number $i \in \mathbb{N}_0$.

Furthermore, we calculate the pair correlation function
\begin{equation}
 g({\bf r}) = \langle \rho_{\text{Rydberg}}({\bf r}) \rho_{\text{Rydberg}}(0)  \rangle,
\end{equation}
where $\rho_{\text{Rydberg}}({\bf r}) d{\bf r}$ denotes the number of Rydberg atoms in a volume element $d{\bf r}$ around the point ${\bf r}$. The first step is
to divide the interval of possible distances between two atoms ($[0;L/2]$ for periodic boundary conditions) into $k$ subintervals of equal length. Calculating
the ground state $|GS\rangle$ for a given distribution of atoms yields the coefficients $v_{i}$. Now, we consider a single pair of atoms and measure their distance in the current distribution of atoms. This distance lies within one of the aforementioned
subintervals. To the value which is already stored for this subinterval we now add the sum of all $|v_{i}|^2$ that correspond to a basis state in which both of the atoms
of the considered
pair are in the Rydberg state. After repeating this procedure for every possible pair of atoms we start over by generating a new random distribution. The
cumulative sum of all samples treated this way then gives the total correlation function for a given set of parameters.


\section{Results and discussion}
\label{sec:results}

Let us first discuss the simplest case of a noninteracting system, $C=0$. In this case, the problem is exactly solvable and the number of Rydberg atoms is given by

\begin{equation}
M_0 = \frac{N}{2} \left(1 + \frac{1}{\sqrt{1 + \Omega^2 / \Delta^2}} \right) \, .
\label{eq:nonintmag}
\end{equation}

In the noninteracting limit, this value is independent of the positions ${\bf r}_i$ of the atoms. Therefore, the histogram $P(M)$ becomes trivial, $P(M) = \delta(M - M_0)$. The density distribution $\rho_0 = \langle \rho_{\text{Rydberg}}({\bf r}) \rangle$, averaged over many realizations, is uniform, and the pair correlation function $g({\bf r}) = \rho_0^2$ is constant.

\begin{figure}[bt]
 \centering
 \begin{minipage}{\columnwidth}
 \includegraphics[width=0.33\columnwidth,keepaspectratio=true]{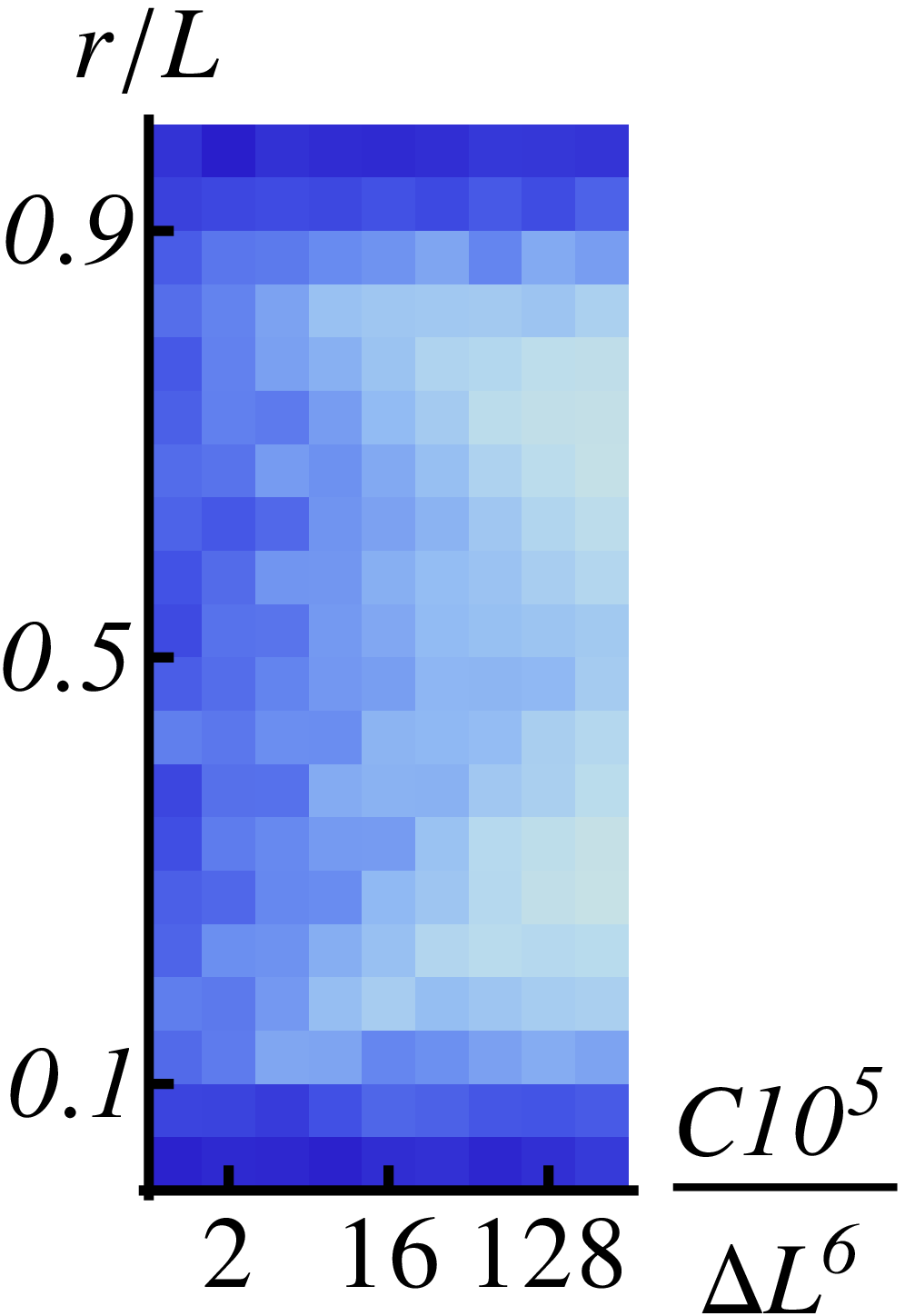}
 \hfill
 \includegraphics[width=0.55\columnwidth,keepaspectratio=true]{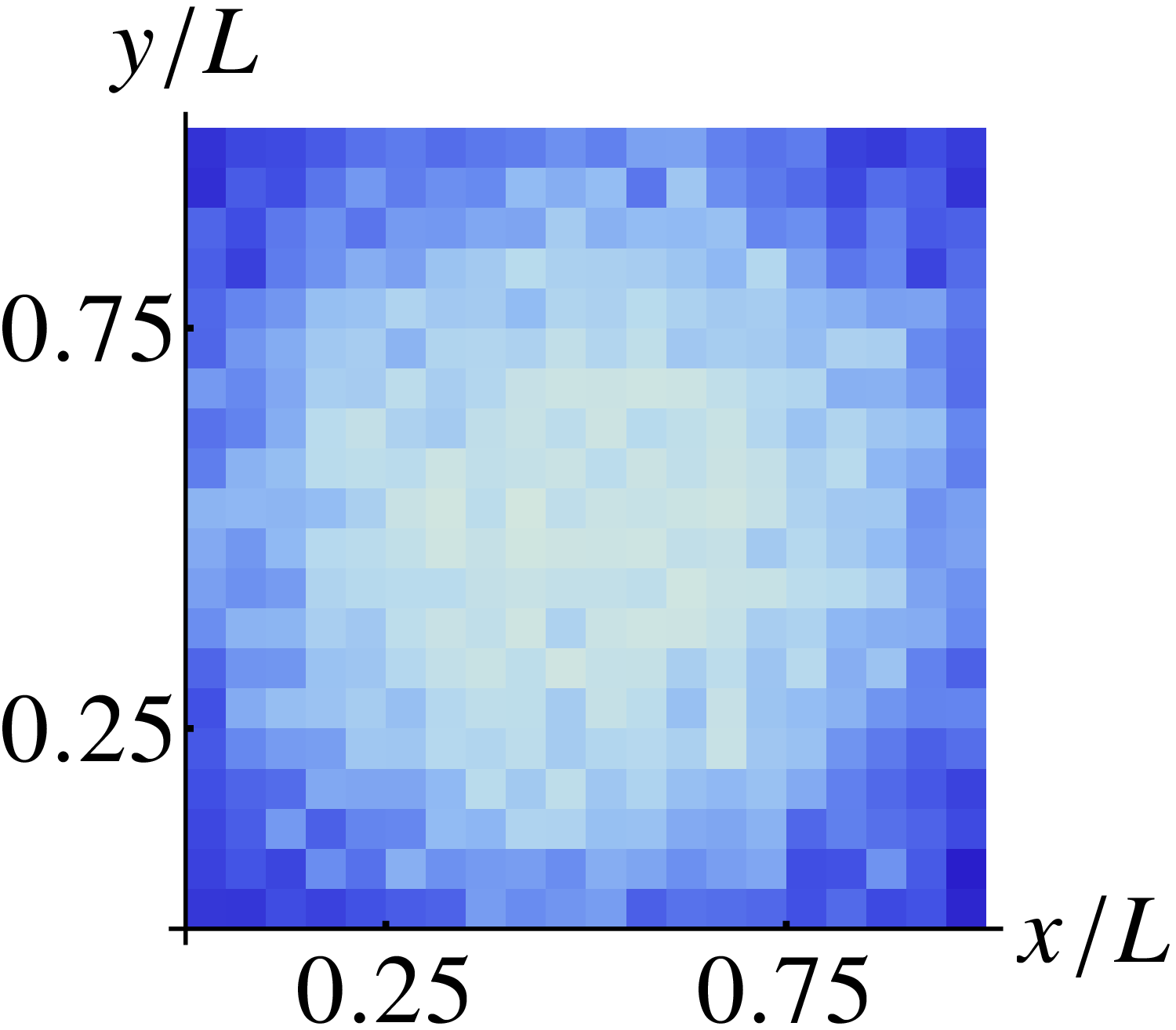}
 \end{minipage}
 \caption{Density distribution of Rydberg atoms, darker color indicates higher density.
 \emph{Left panel:} 1D density as a function of interaction strength. The system boundaries are preferred. Strong interactions produce a single peak in the center.
 \emph{Parameters:} $\Omega / \Delta=0.1$ and $N=M^{*}=6$ atoms (no cutoff, exact diagonalization).
 \emph{Right panel:} 2D projection of a 3D system with $N=M^{*}=6$, $\Omega/\Delta=4$ and $C=10$ in a cube with open boundaries. There is a low
	  density at the center of the volume while it is high in the corners.}
 \label{fig:densitydistribution3Dblue}
\end{figure}

The situation changes drastically for any nonzero $C$. Figure~\ref{fig:densitydistribution3Dblue} shows the data for the density distribution $\langle \rho_{\text{Rydberg}}({\bf r}) \rangle$ in 1D and 3D
systems with open boundaries. While for weak interactions the Rydberg atoms tend to populate the boundaries, in the case of strong interactions a sizeable
fraction of the atoms is redistributed towards the system's bulk, indicating that long-range order is established in the system. As we are dealing with a system
with long-range interactions, the numerical complexity is determined by the number of Rydberg atoms and not by the precise geometry of the system. On the other hand, going to higher dimensions at fixed atom number $N$ increases the surface/bulk
ratio and thus makes the detection of long-range order more cumbersome. Therefore we shall concentrate on (quasi)-1D systems from now on. Nonetheless, we have
conducted a number of simulations on 3D systems and found comparable results, see e.g., Fig.~\ref{fig:densitydistribution3Dblue} for the density profile
of a 3D system. We also would like to remark that an experimentally realizable cigar-shaped confining potential (with a typical radius smaller than
$R_B$) can be very well be approximated by the (quasi)-1D geometry considered here.

\begin{figure}[tb]
 \centering
 \begin{overpic}[width=1\columnwidth,keepaspectratio=true]{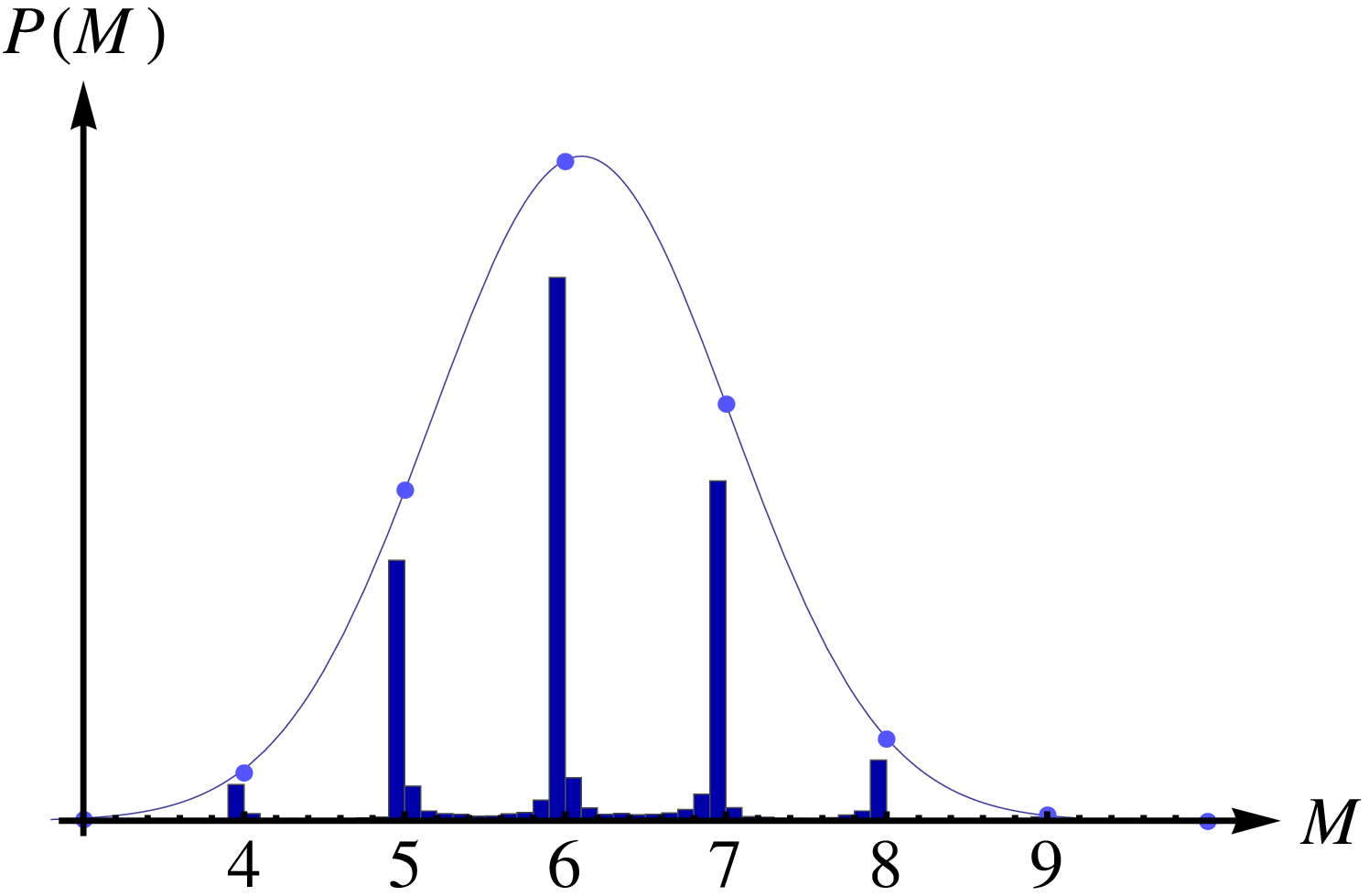}
 \put(50,30){\includegraphics[width=0.5\columnwidth,keepaspectratio=true]{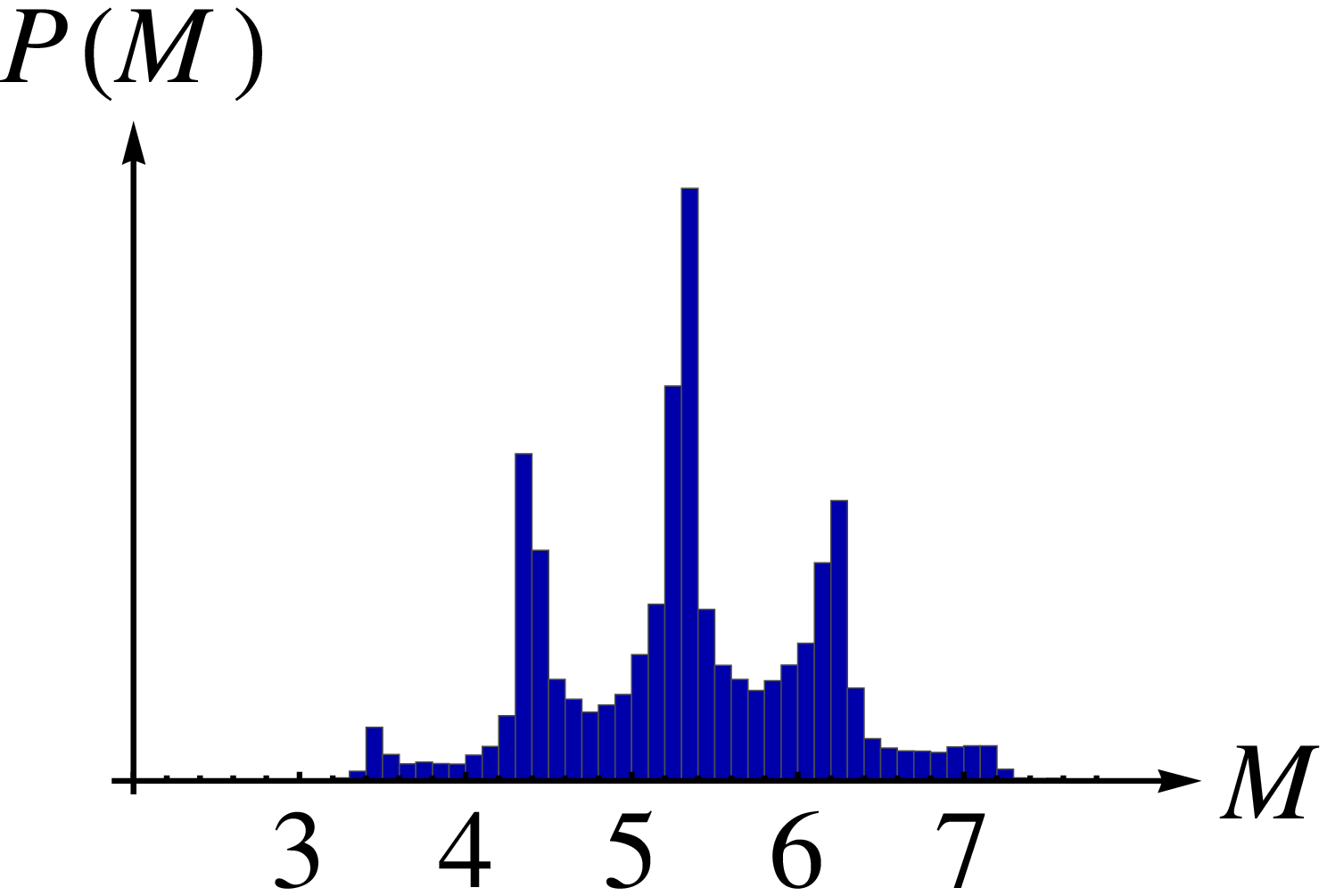}}
 \end{overpic}
 \caption{\emph{Main figure: }Histogram of the number of Rydberg atoms with Gaussian fit. The data points shown are the cumulative sum of all bins in the interval
	  $\lbrack n-0.5,n+0.5 \rbrack$. The
	  features of the plot are explained in the text. Parameters: $\Omega / \Delta =0.1$, $C / (\Delta L^6)=4 \cdot 10^{-7}$ with $N=10=M^*$ atoms (which
	  corresponds to no cut-off). \emph{Inset:} Histogram for $\Omega / \Delta =0.64$, $C / (\Delta L^6)= 5 \cdot 10^{-7}$ with $N=10=M^{*}$ atoms. }
 \label{fig:maghisto1Dandfit}
\end{figure}

Figure~\ref{fig:maghisto1Dandfit} shows a typical result for the histogram $P(M)$. For $\Omega = 0$, the number of Rydberg atoms commutes with the Hamiltonian, so the
ground state for a given set of position $\{ {\bf r}_i \}$ has an integer expectation value $M$ of Rydberg atoms. The distribution $P(M)$, generated by considering
all arrangements $\{ {\bf r}_i \}$, thus becomes a series of discrete, weighted peaks at integer values $M$. For small $0 < \Omega \ll \Delta$, these peaks broaden
up, but the distribution remains more or less discrete. Increasing $\Omega$ leads to further broadening until eventually a continuous distribution is approached, see inset
of Fig.~\ref{fig:maghisto1Dandfit}.

\begin{figure}
 \includegraphics[width=0.49\columnwidth]{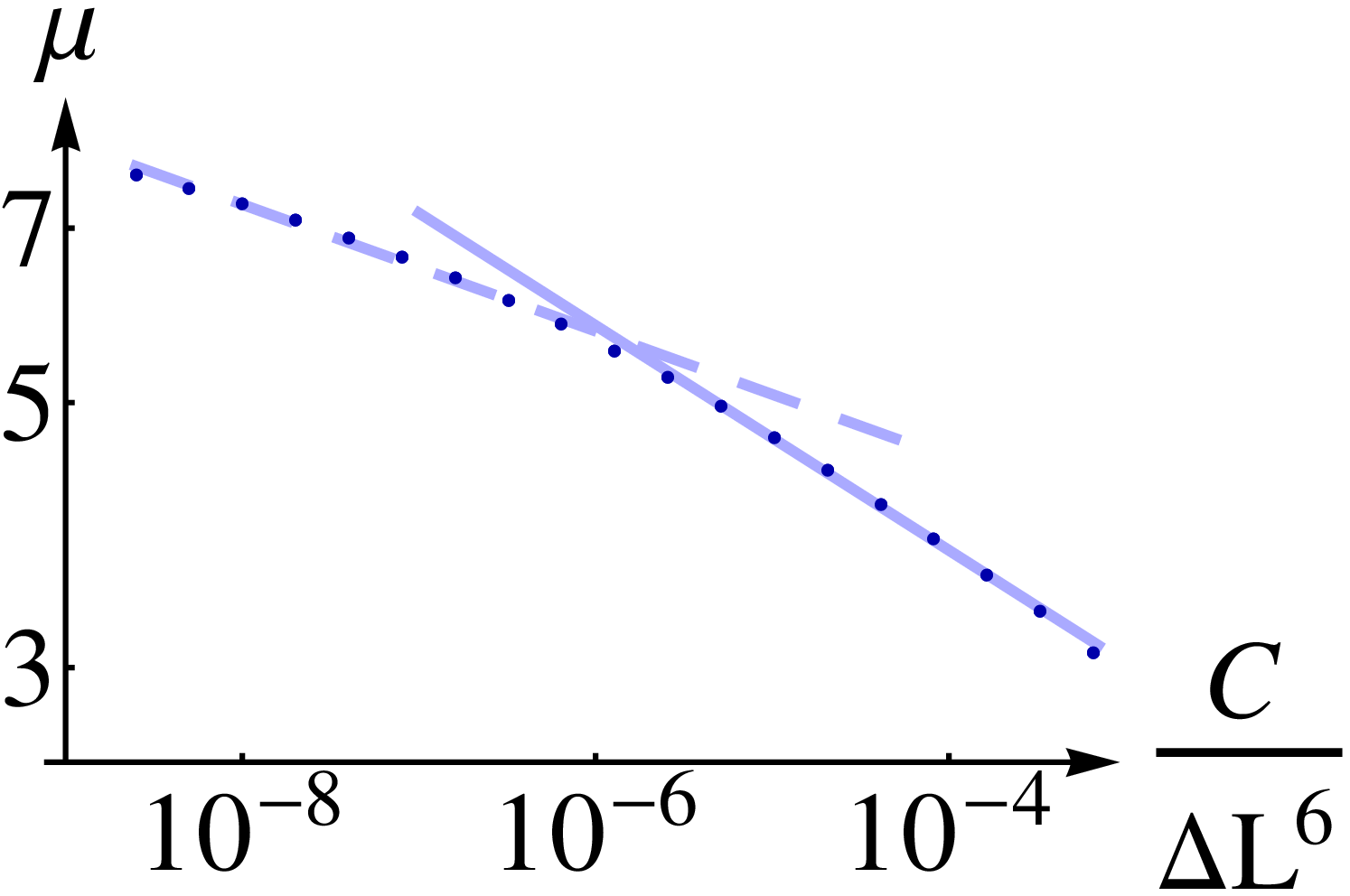}
 \hfill
 \includegraphics[width=0.49\columnwidth]{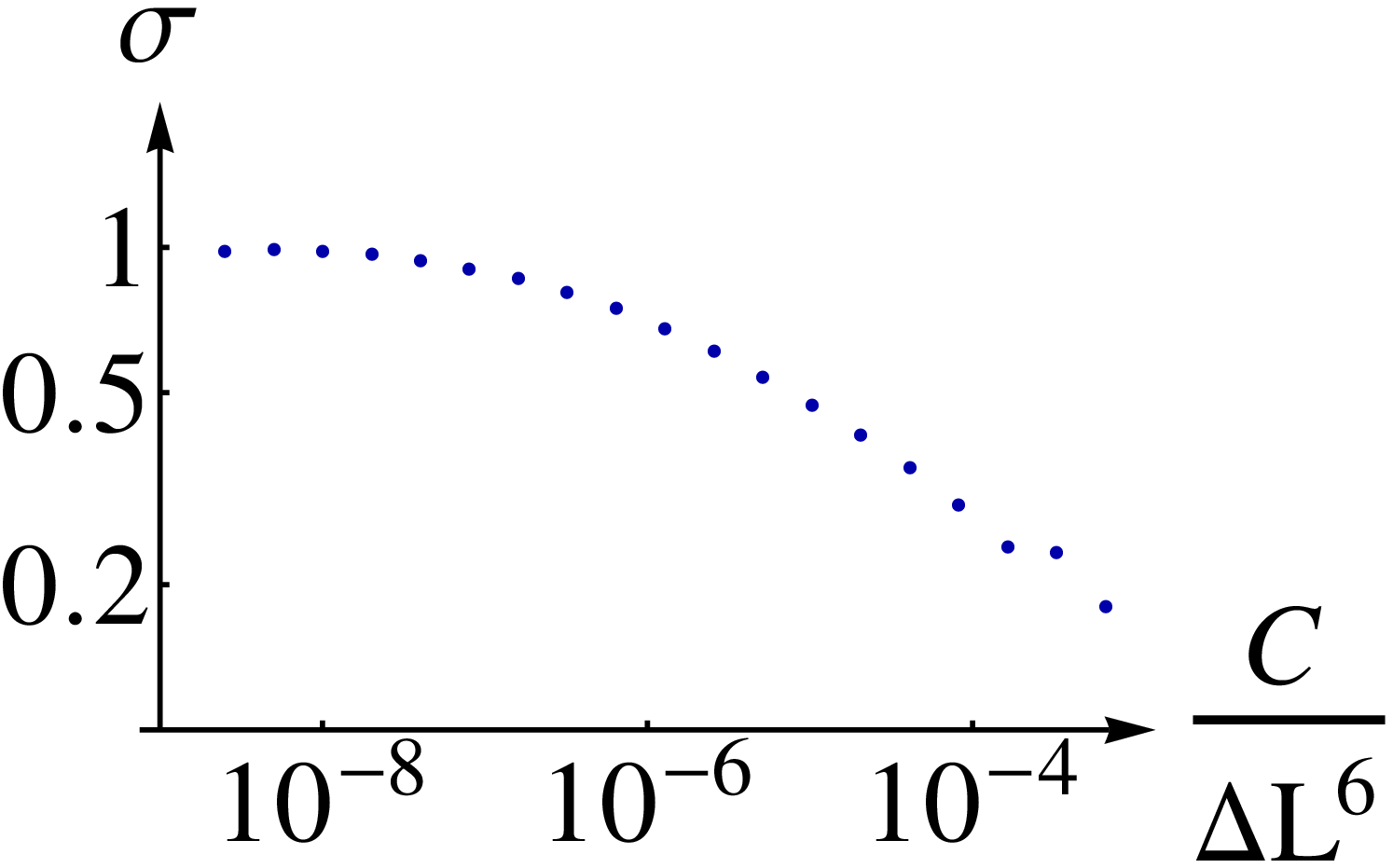}
\caption{Average number of Rydberg atoms $\mu$ (left) and variance $\sigma$ (right) in dependence on $C$ (double-log-plots). The straight lines in the average number
	  of Rydberg atoms plot are guides to the eye only. The parameters for all three plots are $\Omega / \Delta=0.2$ with $N = 10 = M^{*}$ atoms. The resulting
	  Mandel parameter is displayed in Fig.~\ref{fig:mandel}.}
\label{fig:meansigma}
\end{figure}

Surprisingly, in most cases the envelope of the histogram can fairly well be fitted by a Gaussian, as opposed to the Poissonian used, e.g., in
Ref.~[\onlinecite{PhysRevLett.95.253002}]. Nonetheless, the distribution function's higher order cumulants are not exactly zero and depend on $C$, which
indicates a slight deviation from the
Gaussian distribution. The mean $\mu$ and the variance $\sigma$ for a
series of simulations are plotted in Fig.~\ref{fig:meansigma}. Very interestingly, the $\mu(C)$ dependence is given by a power law with an exponent which
changes abruptly from $\approx -0.05$ to $\approx -0.1$ at around $C/(\Delta L^6) \approx 10^{-6}$, where $L$ is the system length. We find this change to be a
harbinger of a phase transition
in the system.
Both the mean and the width of the distribution function decrease as a function of $C$ because a stronger repulsion  increases the blockade radius $R_B$. This
qualitative behavior is independent of the value of $\Omega$.


If one plots the mean and the variance as functions of $\Omega$ the general behavior of the mean $\mu(\Omega)$ is qualitatively the same as for the noninteracting
case [$\mu(\Omega) = \mu(\Omega, C=0)$; cf. Eq. \eqref{eq:nonintmag}] for $\Omega < \Omega^{*}$, where $\Omega^{*} = \text{max}(C/L^6,\Delta)$ is the largest energy scale 
in the system. For  $\Omega > \Omega^{*}$ the spin flip term dominates the Hamiltonian \eqref{eq:hamiltonian} and $M$ tends to $N/2$. The only effect of $C$ is the 
change in the overall amplitude of the curve.
\begin{figure}
  \includegraphics[width=0.7\columnwidth]{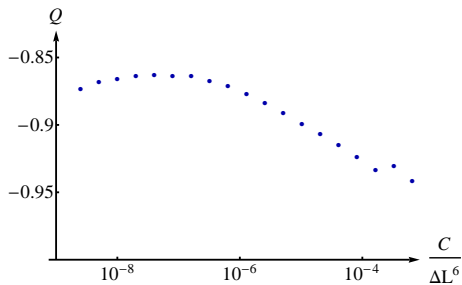}
\caption{Mandel $Q$ parameter in dependence on the interaction strength $C$. The data points are calculated using the data from Fig.~\ref{fig:meansigma}.}
\label{fig:mandel}
\end{figure}

To quantify the deviation of $P(M)$ from a Poissonian distribution we use the Mandel $Q$ parameter which is defined as\cite{Mandel:79}
\begin{equation}
 Q = \frac{\langle M^2 \rangle - \langle M \rangle^2}{\langle M \rangle} - 1.
\end{equation}
It is plotted in Fig.~\ref{fig:mandel} for the same series of simulations. Being negative throughout, it indicates that the $P(M)$ distribution is
sub-Poissonian.\cite{Ates2006, 0953-4075-39-23-007} As realized in Ref.~[\onlinecite{PhysRevLett.95.253002}] this points towards an efficient Rydberg blockade.

Another interesting quantity is the pair correlation function $g(r)$, which is defined as the probability to find a Rydberg atom at a distance
$r$ from another Rydberg atom. In order to exclude boundary effects, we now switch to periodic boundary conditions. To be able to handle computations with
larger numbers of atoms we change the truncation procedure to one in which we consider only a fixed number of basis states. This enables us to approach
$N\approx 30$. This set of states is chosen to be the one with the smallest diagonal elements in the Hamiltonian matrix. We thoroughly investigated the
effect of the truncation by considering the same system with different numbers of contributing states. We find that the result is almost independent of
the number of states as long as it exceeds a certain threshold. All plots displayed in this paper meet this requirement.

\begin{figure}
  \includegraphics[width=0.45\columnwidth]{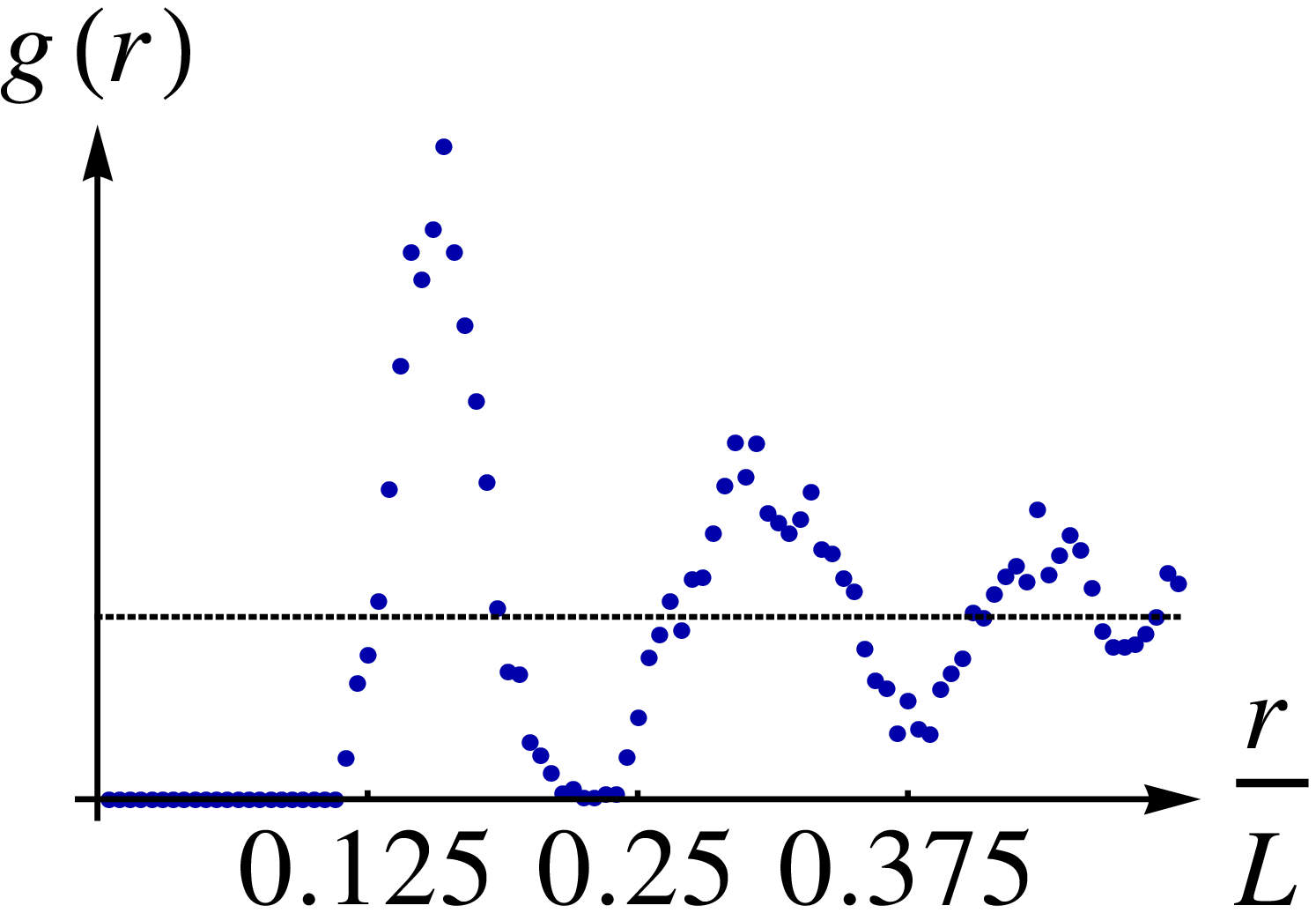}
  \hfill
  \includegraphics[width=0.45\columnwidth]{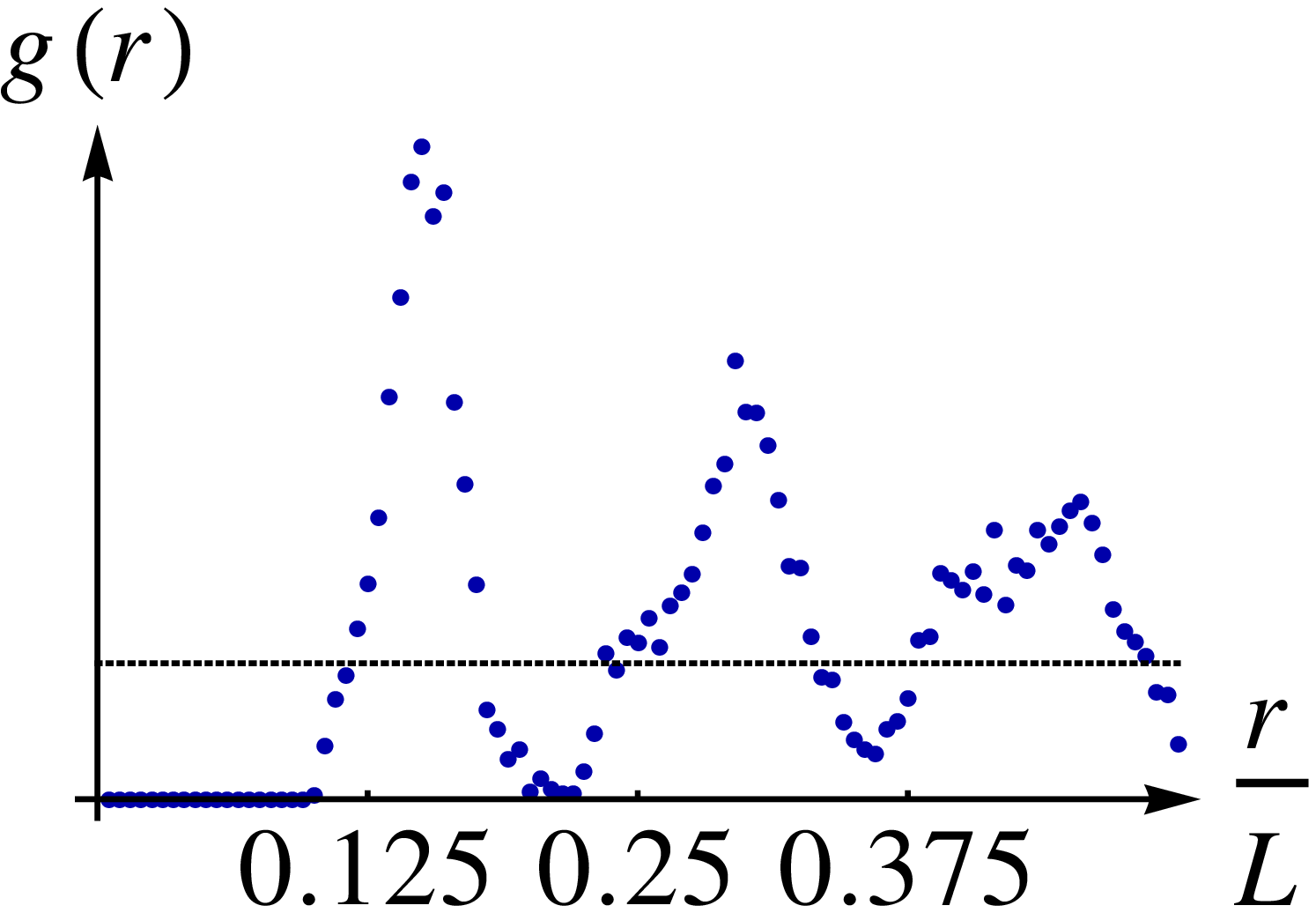}
\caption{\emph{Left:} Correlation function with adjustment of the system length in such a way that main and secondary peaks coincide, $C/(\Delta L^6) = 7.5 \cdot 10^{-6}$.
	  \emph{Right:} Correlation function as a function of distance. The aliasing can be seen in the shoulders of the second and third maximum, $C/(\Delta L^6) = 4 \cdot 10^{-6}$.
	  For both correlation functions $N=30$ and $\Omega/\Delta = 25$.}
\label{fig:aliasing}
\end{figure}

\begin{figure}
  \includegraphics[width=0.7\columnwidth]{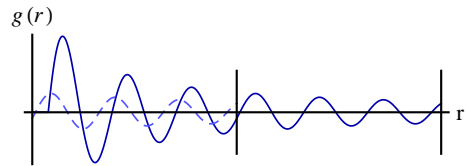}
\caption{Schematic illustration of the effect shown in Fig.~\ref{fig:aliasing}. The dashed line is part of the correlation function in the right interval shifted
	  by the length of the interval. The sum of dashed and solid curve in the left interval is a curve with main peaks that have secondary peaks on their shoulders.
	  The secondary peak on the shoulder of the first peak is suppressed because of the blockade phenomenon.}
\label{fig:schema}
\end{figure}

\begin{figure}[tb]
 \centering
 \includegraphics[width=0.9\columnwidth,keepaspectratio=true]{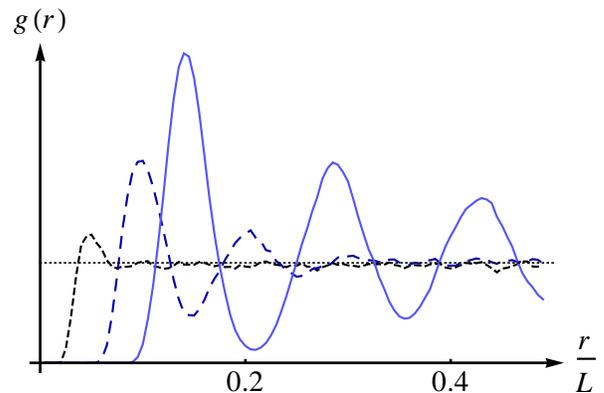}
 \caption{Correlation function of 1D samples measured in arbitrary units. The curves correspond to $C / \Delta L^6 = 4 \cdot 10^{-9}$ (dotted),
	  $C / \Delta L^6=4 \cdot 10^{-7}$ (dashed) and $C/ \Delta L^6=4 \cdot 10^{-6}$ (solid) where all other parameters remained the same:
	  $\Omega / \Delta = 1$ and $N = 25$ with 300 basis states. The curves indicate that there is a critical value for $C / \Delta L^6$ between the two larger
	  values given above.}
 \label{fig:1dimcorrfunction}
\end{figure}

A typical correlation function is shown in Fig.~\ref{fig:aliasing} and is qualitatively comparable to the ones shown in Ref.~[\onlinecite{2012arXiv1202.2012A}]. On
the right panel one can see additional smaller peaks, which originate from the periodic boundary conditions, on the left side of two principal peaks. This feature, which
is known as ``aliasing'', arises in the following way: in the plots shown in Fig.~\ref{fig:aliasing} (and in any other plot of a correlation function in this work) our domain
of definition is the interval $[0,L/2]$ since $L/2$ is the maximum distance between two atoms in a 1D system with periodic boundary conditions. Now we could
extrapolate this correlation function for larger distances. Since we are not able to resolve those the correlation function is mapped onto the given
interval periodically. So the additional peaks appearing here are basically the 5th and 6th order peaks of the correlation function. Fig.~\ref{fig:schema} illustrates
this behaviour, which is also known from the numerical realization of Fourier transforms with finite frequency cutoff.\cite{Press:2007:NRE:1403886} This effect can
be hidden when one uses such parameter values at
which $L$ and $R_B$ are commensurate, see the left panel of Fig.~\ref{fig:aliasing}. Using this control prescription we have calculated $g(r)$ for a very large number
of samples for a number of different interaction strengths, see Fig.~\ref{fig:1dimcorrfunction}.

\begin{figure}
  \begin{overpic}[width=\columnwidth]{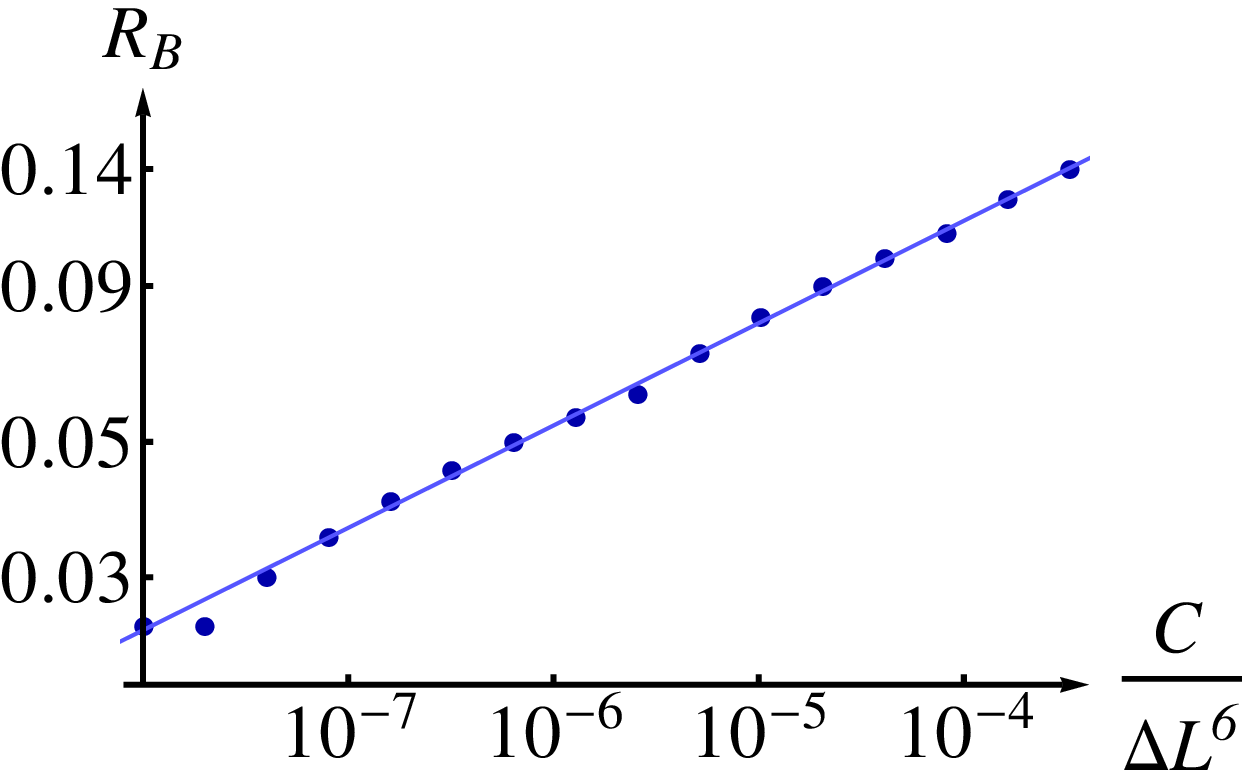}
    \put(17,35){\includegraphics[width=0.42\columnwidth]{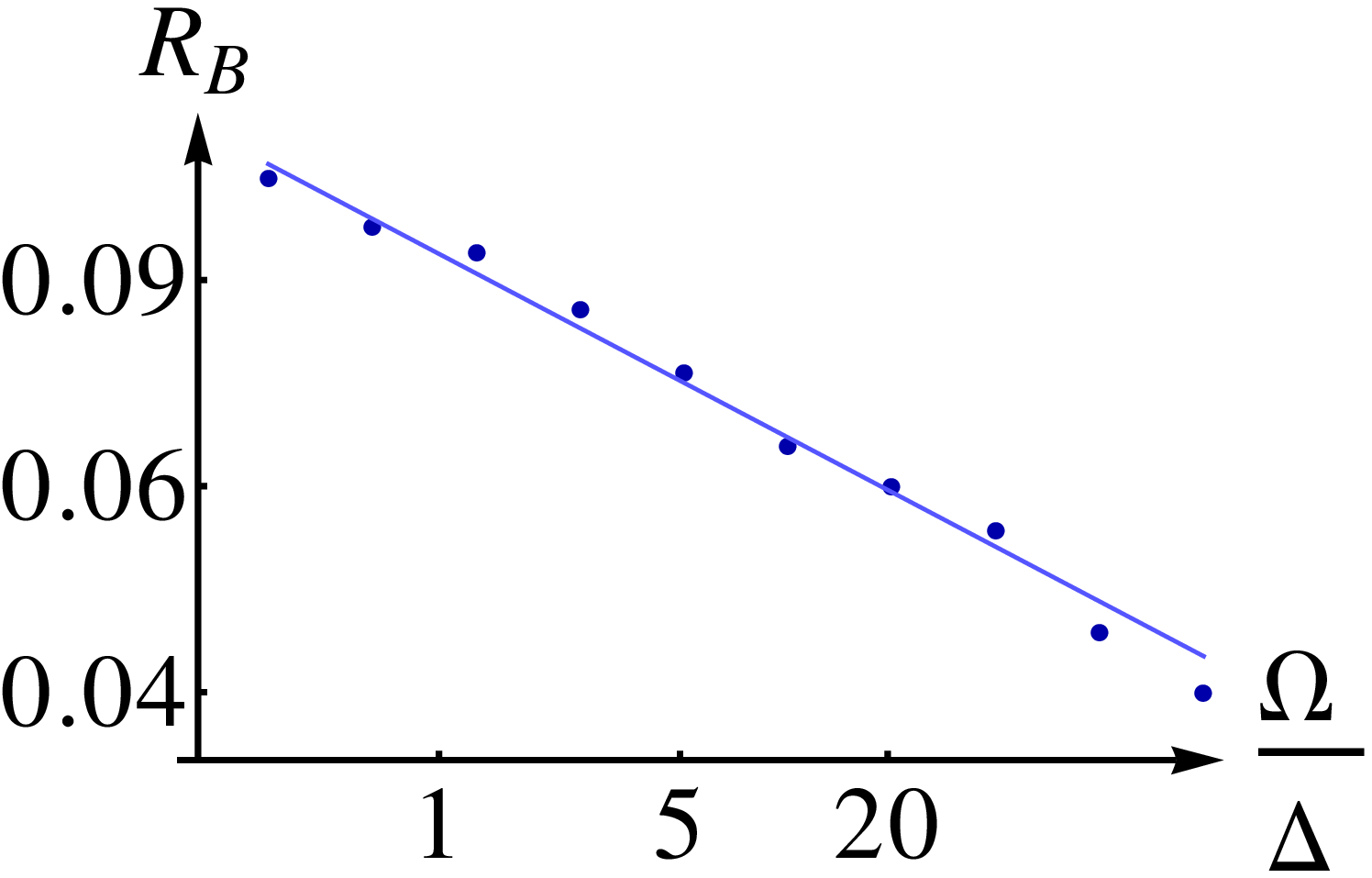}}
  \end{overpic}
  \caption{\emph{Main graph:} $R_B$ as a function of $C$ (double-log-plot). The data points clearly show a power law. The parameters are:
	  $\Omega / \Delta=0.1$ with $N=8=M^{*}$ atoms. The fit is
	  discussed in the text. \emph{Inset:} $R_B$ as a function of $\Omega$ (double-log-plot). The parameters are: $C/(\Delta L^6)=10^{-5}$ with $N=7=M^{*}$ atoms.}
\label{fig:blockadeC}
\end{figure}

As expected, one can immediately identify the blockade radius $R_B$ as the position of the first maximum. It is the distance between two atoms up to which
it is disadvantageous to excite both of them to the Rydberg state. Obviously, it is nonzero for all interaction strengths $C$. The actual $R_B(C)$ dependence
is very interesting and is plotted in Fig.~\ref{fig:blockadeC}. Very naturally, $R_B$ grows with interaction strength as
\begin{equation}
 R_B \approx (C/\Delta L^6)^\gamma \, ,
\end{equation}
where $\gamma \approx 1/6 \pm 1\%$. This was also discussed in
Refs.~[\onlinecite{PhysRevLett.99.163601,2011PhRvL.107o3001S,PhysRevLett.105.193603,PhysRevLett.107.103001}].
\begin{figure}
  \begin{overpic}[width=\columnwidth]{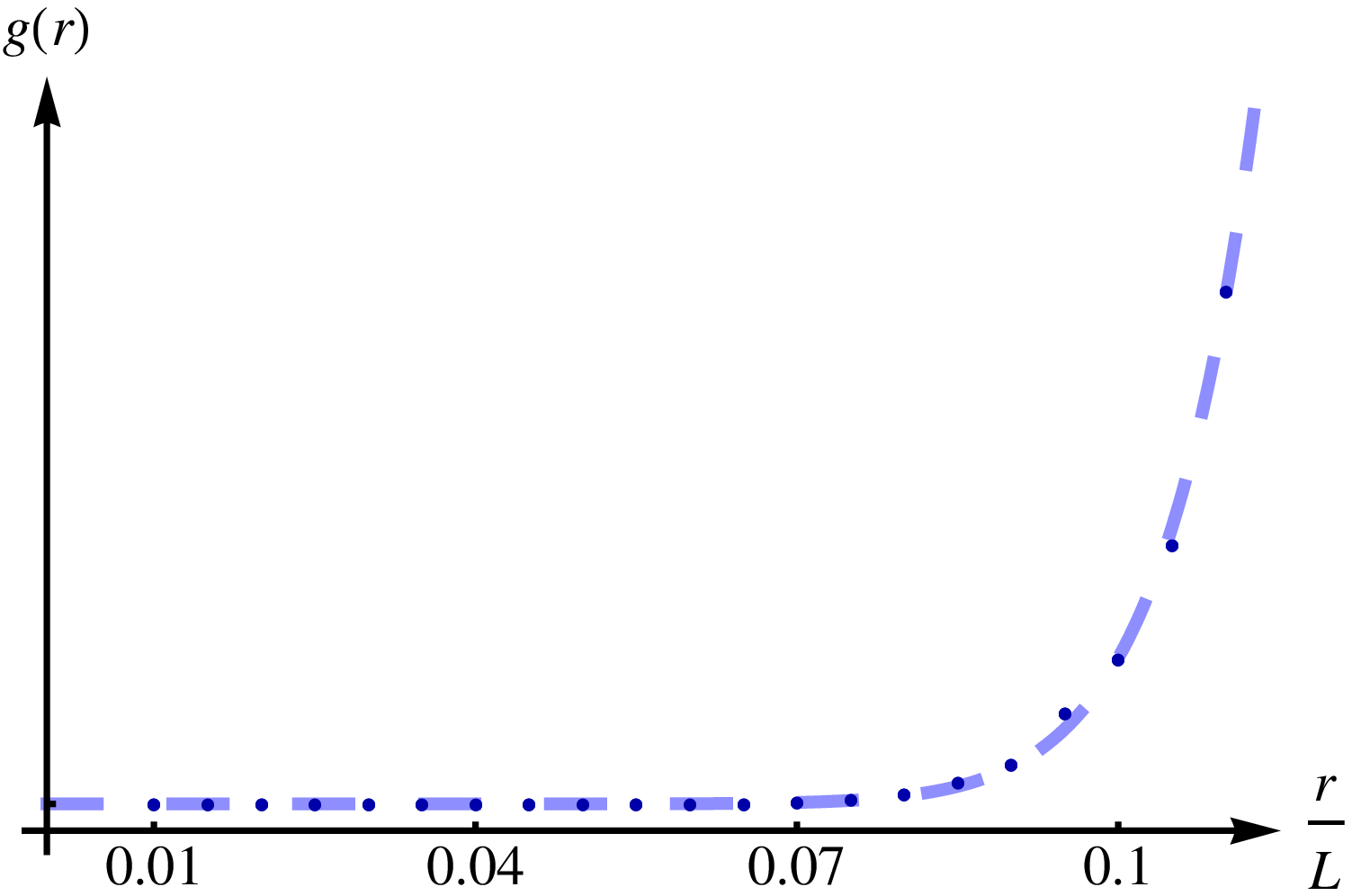}
    \put(6,13){\includegraphics[width=0.7\columnwidth]{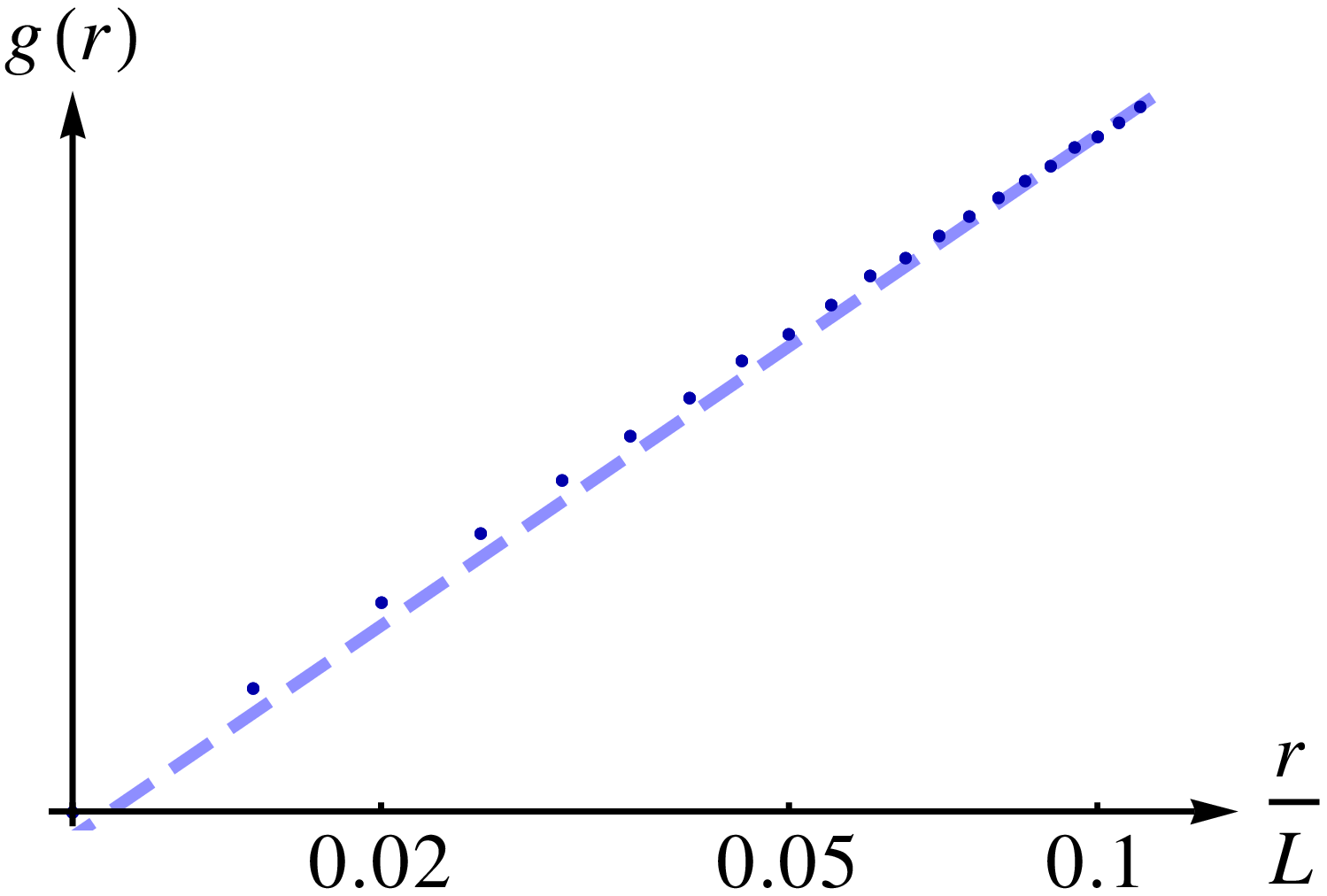}}
  \end{overpic}
\caption{\emph{Main graph:} Correlation function as a function of distance. The dashed line represents a fit explained in the text.
	  \emph{Inset:} The same plot as in the main plot shown on a logarithmic scale. The parameters of the plot are:
	  $\Omega / \Delta=1$, $C/(\Delta L^6)=2\cdot10^{-5}$ with $N=6=M^{*}$ atoms.}
\label{fig:corrfoot}
\end{figure}
Furthermore, we find the asymptotic behavior of the
correlation function for $r \ll R_B$ to be highly universal (see Fig.~\ref{fig:corrfoot}). It is given by
\begin{equation}
 g(r)  \propto r^{12}, \qquad r \ll R_B
\end{equation}
which is in contrast to the step-like behavior found in Ref.~[\onlinecite{2012arXiv1202.2012A}].
Beyond the point $r=R_B$ the qualitative shape of $g(r)$ depends strongly on $C$. While for weak interactions, $C/(\Delta L^6) < 4 \cdot 10^{-9}$,
no additional peaks can be seen, long-range order emerges for strong interactions $C/(\Delta L^6) >  10^{-6}$, and manifests itself as a series of equidistant peaks.
In the former case, the Rydberg atoms remain in a gas phase, whereas the latter situation might by described as the Rydberg crystal proposed in
Refs.~[\onlinecite{PhysRevLett.104.043002,1367-2630-12-10-103044}]. In an ideal crystal, the additional peaks would be sharp. In our calculations this cannot be
achieved due to finite-size effects. From our simulations we estimate the dimensionless critical parameter for the transition as
$C_\text{crit}/(\Delta L^6) \approx 5 \cdot 10^{-7}$.

In fact, one can recognize this critical value for the phase transition already in the simple statistical parameter of the $P(M)$ histograms. Indeed, not only
the mean value $\mu$, but also the variance and Mandel parameter $Q$ remarkably change their behavior near the critical value $C_\text{crit}$. For instance, $Q$ is
roughly constant for $C<C_\text{crit}$ but starts to decrease just below $C_\text{crit}$, see Fig.~\ref{fig:mandel}. The same happens to the variance. The behavior of
$\mu$ is less prominent but still clearly detectable. Similar behavior can be seen in higher cumulants of $P(M)$. However, the larger statistical errors might make
them less useful in practical applications.




\section{Conclusions}
\label{sec:sumandcon}

Using exact numerical diagonalization and approximative descendants of this method we have investigated the Rydberg crystallization phenomenon in ultracold gases.
We have estimated the critical interaction strength by two different techniques. Both pair correlation
function and simpler statistical data show signatures of this phase transition. The big advantage of using mean and variance of the histogram for the number of
excited atoms measured in a long series of identical experiments is its good experimental accessibility. In this way we have developed a purely statistical
method of detecting Rydberg crystallization. We hope that this procedure can soon be implemented in state-of-the-art experiments in order to unambiguously
identify the Rydberg crystallization phenomenon without complicated spectroscopic techniques.

Furthermore, the presented details of the pair correlation function might be useful for the continuous experimental efforts in spatial imaging of Rydberg aggregates
in the spirit of Refs.~[\onlinecite{PhysRevLett.107.103001, 2012PhRvL.108a3002G}]. This technique is not only able to yield very precise value of the blockade radius,
but has also proven to be a reliable source for the estimation of the critical parameters for Rydberg crystallization.

%
%


\acknowledgments
The authors would like to thank G.~G\"unther, M.~Weidem\"uller and S.~Whitlock for many interesting discussions.
AK is supported by the DFG Grant No. KO-2235/3-1, CQD and `Enable fund' of the University of Heidelberg.


\end{document}